\documentclass[11pt,a4paper]{article}
\usepackage{amsmath}
\usepackage{amssymb}
\usepackage{amsthm}
\usepackage{times}
\usepackage{a4}
\usepackage[hang,stable]{footmisc}

\usepackage{bbm}
\usepackage{color}
\usepackage{graphicx}
\usepackage{subfig}
\usepackage{units}
\usepackage{caption}
\newcommand{\e}{\mathrm{e}}
\newcommand{\abs}[1]{\vert #1 \vert}
\newcommand{\D}{\mathrm{d}}
\newcommand{\bleq}{\mathrel{\phantom{=}}}

\newcommand{\reals}{\mathbb{R}}
\newcommand{\complex}{\mathbb{C}}

\begin{document}
\ifx\href\undefined\else\hypersetup{linktocpage=true}\fi

\newtheorem{theorem}{Theorem}
\newtheorem{proposition}{Proposition}
\newtheorem{Quote}{Einstein Quote}
\theoremstyle{definition}
\newtheorem{Requirement}{Requirement}
\newtheorem{Remark}{Remark}

\title{Gravitationally induced inhibitions of dispersion\\ 
       according to the Schr\"odinger-Newton Equation}
\author{Domenico Giulini and Andr\'e Gro{\ss}ardt       \\
        Center of Applied Space Technology and Microgravity\\
        University of Bremen             \\
        Am Fallturm 1                    \\
        D-28359 Bremen, Germany          \\
        and                              \\
        Institute for Theoretical Physics\\
        Leibniz University Hannover      \\
        Appelstrasse 2                   \\
        D-30167 Hannover, Germany}

\date{}

\maketitle

\begin{abstract}
\noindent
We re-consider the time dependent Schr\"odinger-Newton equation as a 
model for the self-gravitational interaction of a quantum 
system. We numerically locate the onset of gravitationally 
induced inhibitions of dispersion of Gaussian wave packets 
and find them to occur at mass values more than 6 orders of 
magnitude higher than reported by Salzman and Carlip
\cite{Salzman-Carlip:2006,Carlip:2008}, namely at about 
$10^{10}\,\mathrm{u}$. This fits much better to simple analytical 
estimates but unfortunately also questions the experimental 
realisability of the proposed laboratory test of quantum gravity 
in the foreseeable future, not just because of large masses, 
but also because of the need to provide sufficiently long 
coherence times.  
\end{abstract}

%\begin{small}
%\setcounter{tocdepth}{3}
%\tableofcontents
%\end{small}
%\newpage

\section{Introduction}
In this paper we investigate the time dependent 
Schr\"odinger-Newton equation (henceforth abbreviated 
SN-equation):
\begin{equation}
\label{eq:SchroedingerNewton}
i\hbar\partial_t\Psi(t,\vec x)=
\left(
-\frac{\hbar^2}{2m}\Delta-
Gm^2\int\frac{\vert\Psi(t,\vec y)\vert^2}%
{\Vert\vec x-\vec y\Vert}\,d^3y
\right)
\Psi(t,\vec x)\,.
\end{equation}
It can be thought of as ordinary time-dependent 
Schr\"odinger equation,
\begin{equation}
\label{eq:Schroedinger}
i\hbar\partial_t\Psi(t,\vec x)=
\left(
-\frac{\hbar^2}{2m}\Delta+m\Phi(t,\vec x)
\right)
\Psi(t,\vec x)\,,
\end{equation}
with a Newtonian gravitational potential $\Phi$ sourced by 
a mass distribution associated to $\Psi$ itself, that is 
$m\vert\Psi\vert^2$, so that 
\begin{equation}
\label{eq:Newton}
\Delta\Phi(t,\vec x)=4\pi G\,m\,\vert\Psi(t,\vec x)\vert^2\,.
\end{equation}

The re-coupling of $\Psi$ via (\ref{eq:Newton}) into 
(\ref{eq:Schroedinger}) results in the non-linear and non-local 
self-interaction seen in (\ref{eq:SchroedingerNewton}). Physical 
intuition suggests this self-interaction to cause a slow down of 
dispersion which should be the more pronounced
%the smaller the widths (support) of $\Psi$ and
the higher the value of $m$ is chosen.

In \cite{Salzman-Carlip:2006,Carlip:2008} it was surprisingly found 
on the basis of numerical computations that Gaussian wave packets of 
width $a\approx 0.5\,\mu\mathrm{m}$ start shrinking in width if the 
mass parameter exceeds $m\approx 1600\,\mathrm{u}$. This is surprising 
in view of the fact that simple analytical estimates suggest this 
type of behaviour to occur much later, namely at the said initial width 
for mass values between around $10^{10}\,\mathrm{u}$. The authors of  
\cite{Salzman-Carlip:2006,Carlip:2008} observed this discrepancy and 
encouraged a check of their results, which we performed. The result is, 
that our numerical studies now fully confirm the high mass values so that 
no discrepancy between numerical and simple analytical estimates seem 
to occur.  

The special attention to typical width dimension of 
$a\approx 0.5\,\mu\mathrm{m}$ stems from actual molecular 
interferometry experiments in which the wave nature of 
complex molecules has been demonstrated; see, e.\,g., 
\cite{Arndt.Hornberger.Zeilinger:2005} for an overview. 
The molecules used in \cite{Hackermueller.etal:2003} were 
tetraphenylporphyrin $C_{44}H_{30}N_4$ with mass 
$614\,\mathrm{u}$ and the more complex and more massive 
fluorofullerene $C_{60}F_{48}$ composed of 108 atoms 
and of total mass $1632\,\mathrm{u}$. The grating period
and grating thickness were about $991\,\mathrm{nm}$ 
and $500\,\mathrm{nm}$ respectively.

For further discussion, we now collect some background material 
concerning the SN-equation~(\ref{eq:SchroedingerNewton}).

\section{Basic features of the 
Schr\"odinger-Newton Equation}
Equation (\ref{eq:SchroedingerNewton}) can be derived from the 
following action (we write $\dot\Psi$ instead of 
$\partial_t\Psi$ and $\Psi^*$ for the complex conjugate of $\Psi$):
\begin{alignat}{1}
\mathcal{S}[\Psi,\Psi^*]=\int \D t
\biggl\{&\frac{i\hbar}{2}\int \D^3x
\Bigl(\Psi^*(t,\vec x)\dot\Psi(t,\vec x)
-\Psi(t,\vec x){\dot\Psi}^*(t,\vec x)\Bigr)\nonumber\\
-\,&\frac{\hbar^2}{2m}\int \D^3x
\bigl(\vec\nabla\Psi(t,\vec x)\bigr)\cdot
\bigl(\vec\nabla\Psi^*(t,\vec x)\bigr)\nonumber\\
\label{eq:NonLocalActionForSN}
+\,&\frac{Gm^2}{2}\iint \D^3x\,\D^3y\,
\frac{\vert\Psi(t,\vec x)\vert^2\,\vert\Psi(t,\vec y)\vert^2}%
{\Vert\vec x-\vec y\Vert}
\biggr\}\,.
\end{alignat}
Note that $\Psi$ and $\Psi^*$ are to be varied independently.
A local form can be obtained if the Newtonian gravitational 
field $\Phi$ is introduced:

\begin{alignat}{1}
\mathcal{S}[\Psi,\Psi^*,\Phi]=\int \D t\int \D^3x
\biggl\{&\frac{i\hbar}{2}
\Bigl(\Psi^*(t,\vec x)\dot\Psi(t,\vec x)
-\Psi(t,\vec x){\dot\Psi}^*(t,\vec x)\Bigr)\nonumber\\
&-\frac{\hbar^2}{2m}
\bigl(\vec\nabla\Psi(t,\vec x)\bigr)\cdot
\bigl(\vec\nabla\Psi^*(t,\vec x)\bigr)\nonumber\\
&-\frac{1}{8\pi G}\bigl(\vec\nabla\Phi(t,\vec x)\bigr)\cdot
\bigl(\vec\nabla\Phi(t,\vec x)\bigr))\nonumber\\
\label{eq:LocalActionForSN}
&- m\,\vert\Psi(t,\vec x)\vert^2\Phi(t,\vec x)
\biggr\}\,.
\end{alignat}
Variation of (\ref{eq:LocalActionForSN}) with respect to 
$\Phi$ gives (\ref{eq:Newton}). Inserting the solution 
$\Phi=4\pi Gm\Delta^{-1}\vert\Psi\vert^2$ into 
(\ref{eq:LocalActionForSN}) then leads to 
(\ref{eq:NonLocalActionForSN}). 

We usually restrict attention to solutions which are square integrable 
over space, so that solutions of the time dependent SN-equations are 
paths in the Hilbert space $L^2(\reals^3,\D^3x)$. We shall see below 
that time evolution preserves the norm (this follows from the symmetry 
under phase transformations).

Symmetries of (\ref{eq:SchroedingerNewton}) are:
\begin{enumerate}
\item
Constant phase shifts
\begin{equation}
\label{eq:ConstPhaseShifts}
\Psi\mapsto\Psi'\,,\quad
\Psi'(t,\vec x):=\exp(i\alpha)\Psi(t,\vec x)
\end{equation}
with constant $\alpha\in\mathbb{R}$. This symmetry of
the equation of motion is also a symmetry of the action
(\ref{eq:NonLocalActionForSN}). Hence there is a conserved 
Noether current which turns out to be just the same 
functional of $\Psi$ as in ordinary Schr\"odinger theory. 
In particular, the space integral over $\vert\Psi\vert^2$ 
is time independent. 

\item
Proper orthochroneous Galilei Transformations 
\begin{equation}
\label{eq:GalileiTrans-1}
(t,\vec x)\rightarrow\tau_g(t,\vec x):=
(t+b\,,\,\mathbf{R}\cdot\vec x+\vec vt+\vec a)\,,
\end{equation}
where $\mathbf{R}\in\mathrm{SO}(3)$ denotes spatial rotations, 
$\vec v\in\mathbb{R}^3$ boost transformations, and $a\in\mathbb{R}^3$
and $b\in\mathbb{R}$ space and time translations respectively,
and where $g$ collectively stands for $(\mathbf{R},\vec v,\vec a,b)$.
These generate the inhomogeneous (proper orthochronous) Galilei
group which acts via proper ray-representations on wave functions 
as follows:
\begin{equation}
\label{eq:GalileiTrans-2}
\Psi\rightarrow T_g\Psi:=\exp(i\beta_g)\bigl(\Psi\circ \tau_{g^{-1}}\bigr) 
\end{equation}
Note that except for the phase $\exp(i\beta_g)$, this is just the 
transformation of a scalar function on spacetime under the 
diffeomorphism $\tau_g$. The $g$-dependent phase function 
$\beta:\mathbb{R}^4\rightarrow\mathbb{R}$ is determined up to 
spacetime independent additions. A convenient choice is 
(compare~\cite{Giulini:1996})
\begin{equation}
\label{eq:GalileiTrans-3}
\beta_g(t,\vec x)=\frac{m}{\hbar}
\Bigl[
\vec v\cdot(\vec x-\vec a)-\tfrac{1}{2}\vec v^2(t-b)
\Bigr]\,. 
\end{equation}
It is immediate that the self-interaction term in 
(\ref{eq:SchroedingerNewton}) and (\ref{eq:NonLocalActionForSN})
does not affect Galilei invariance of the equation of 
motion. Moreover, (\ref{eq:GalileiTrans-2}) is also a 
symmetry of the action so that 10 conserved quantities 
result.  The expressions for total momentum and angular 
momentum are again the same functionals of $\Psi$ as in 
ordinary Schr\"odinger theory, but that for total energy  
will clearly include the non-linear interaction.
\item
Whereas the proper orthochroneous Galilei transformations 
exclude space and time inversions, the latter are also 
symmetries of the SN-equation. To define them, let 
$\pi$ and $\tau$ be the following diffeomorphisms of 
$\reals^4$:
\begin{subequations}
\label{eq:DefReflectionDiffeos}
\begin{alignat}{3}
\label{eq:DefReflectionDiffeosSpace}
\pi&:\reals^4&&\rightarrow\reals^4\,,\quad
(t,\vec x)&&\mapsto (t,-\vec x)\,,\\
\label{eq:DefReflectionDiffeosTime}
\tau&:\reals^4&&\rightarrow\reals^4\,,\quad
(t,\vec x)&&\mapsto (-t,\vec x)\,.
\end{alignat}
\end{subequations}
The unitary transformations of space inversion and the anti-unitary 
transformation of time inversion (or better ``reversal of motion'') 
on wavefunctions is then defined by 
\begin{subequations}
\label{eq:DefReflectionTrans}
\begin{alignat}{2}
\label{eq:DefReflectionTransSpace}
&P\Psi&&:=\Psi\circ\pi^{-1}\,,\\
\label{eq:DefReflectionTransTime}
&T\Psi&&:=C \circ \Psi\circ\tau^{-1}\,,
\end{alignat}
\end{subequations}
where $C$ denotes complex-conjugation. 
It is easy to check that if $\Psi$ solves the SN-equation
then $P\Psi$ and $T\Psi$ also solve it. 
\item
There is a scaling equivariance between SN-equations of 
different mass parameters $m$. The precise 
statement is this: Let $\mathbb{R}_+$ be the multiplicative 
group of positive real numbers. It acts by scaling 
transformations on spacetime: 
\begin{equation}
\label{eq:ScalingActionSpacetime}
\sigma:\mathbb{R}_+\times\mathbb{R}^4\rightarrow\mathbb{R}^4\,,\quad
\sigma\bigl(\mu\,,\,(t,\vec x)\bigr)=
\sigma_\mu(t,\vec x)=(\mu^at\,,\,\mu^b\vec x)\,,
\end{equation}
where $a,b$ are real numbers characterising the action. 
It also acts on the complex numbers:
\begin{equation}
\label{eq:ScalingActionComplex}
s:\mathbb{R}_+\times\mathbb{C}\rightarrow\mathbb{C}\,,\quad
s(\mu,z)=s_\mu(z)=\mu^c z\,,
\end{equation}
where $c$ is yet another real number. Now, since the group 
acts on $\reals^4$ (via $\sigma$) and $\complex$ (via $s$)
it also acts on  $\reals^4\times\complex$ (via $\sigma\times s$)
and hence on graphs of functions $\Psi:\reals^4\rightarrow\complex$. 
The transform of a function (via $S$) is then defined to be the 
function uniqueley corresponding to the transformed graph: 
\begin{equation} 
\begin{split}
\label{eq:GraphTransform}
\mathrm{graph(S_\mu\Psi)}
:&=\sigma_\mu\times s_\mu\bigl(\mathrm{graph}(\Psi)\bigr)\\
 &=\bigcup_{(t,\vec x)\in\reals^4}\Bigl(\sigma_\mu\bigl(t,\vec x\bigr)\,,
\,s_\mu\bigl(\Psi(t,\vec x)\Bigr)\\
 &=\bigcup_{(t,\vec x)\in\reals^4}\Bigl(\bigl(t,\vec x\bigr)\,,
\,s_\mu\circ\Psi\circ\sigma^{-1}_\mu(t,\vec x)\Bigr)\,,\\
\end{split}
\end{equation}
where in the last step we just used that each $\sigma_\mu$ is a bijection. Hence
\begin{equation}
\label{eq:ScalingActionFunctions-1}
\begin{split}
S:\mathbb{R}_+\times\text{Funct}(\mathbb{R}^4,\mathbb{C})
&\rightarrow\text{Funct}(\mathbb{R}^4,\mathbb{C})\,,\\
S(\mu,\Psi)&=S_\mu(\Psi)=
s_\mu\circ\Psi\circ\sigma^{-1}_\mu\,.\\
\end{split}
\end{equation}
or, in simpler terms, 
\begin{equation}
\label{eq:ScalingActionFunctions-2}
S_\mu\Psi(t,\vec x)=\mu^c\,
\Psi(\mu^{-a}t,\mu^{-b}\vec x)\,.
\end{equation}
As already mentioned, we wish to consider normalised wave 
functions. The scaling (\ref{eq:ScalingActionFunctions-1}) 
preserves normalisation iff $2b+3c=0$, i.\,e. iff $c=-3b/2$. Solutions 
to the SN-equation for all positive mass parameters $m$ 
then correspond to special paths on the unit sphere in Hilbert 
space. We ask under what conditions the action 
(\ref{eq:ScalingActionFunctions-1}) transforms solutions 
$\Psi$ to the SN-equation for mass parameter $m$ to solutions 
$S_\mu\Psi$ for some mass parameter $m_\mu$.  We have
\begin{subequations}
\label{eq:ScalingTrans}
\begin{alignat}{1}
\label{eq:ScalingTrans-a}
\partial_tS_\mu\Psi&=\mu^{-a}S_\mu\partial_t\Psi\\
\label{eq:ScalingTrans-b}
m_\mu^{-1}\Delta S_\mu\Psi&=\mu^{-2b}(m/m_\mu)\,m^{-1}
S_\mu\Delta\Psi\\
\label{eq:ScalingTrans-c}
m_\mu^2U(S_\mu\Psi)S_\mu\Psi&=
(m_\mu/m)^2\,\mu^{-b}\,m^2\,S_\mu\bigl(U(\Psi)\Psi\bigr),
\end{alignat}
\end{subequations}
where $U(\Psi)$ is the potential function
\begin{equation}
\label{eq:PotFunct}
U(\Psi)(t,\vec x)=
\int\frac{\vert\Psi(t,\vec y)\vert^2}{\Vert\vec x-\vec y\Vert}\D^3x\,.
\end{equation}
Denoting the SN-equation symbolically by $SN(m,\Psi)=0$, we have
\begin{equation}
\label{eq:SNSymm-1}
SN(m,\Psi)=0 \; \Rightarrow \; SN(m_\mu,S_\mu\Psi)=0
\end{equation}
iff each term scales with the same factor. According to 
(\ref{eq:ScalingTrans}) this is the case iff 
\begin{equation}
\label{eq:SNSymm-2}
\mu^{-a}=\mu^{-2b}\,(m/m_\mu)=\mu^{-b}\,(m_\mu/m)^2
\end{equation}
i.\,e. iff
\begin{equation}
\label{eq:SNSymm-3}
\frac{m_\mu}{m}=\mu^{-b/3}=\mu^{-a/5}\,.
\end{equation}
First of all this means that $m_\mu$ is a power of $\mu$
times $m$. This power may without loss of generality be taken
to unity, for if $m_\mu/m$ were equal to $\mu^d$ we could 
take $\mu':=\mu^d$ as new and equally good scaling parameter 
and everythig that follows were true for $\mu'$ instead of 
$\mu$. So from $m_\mu/m=\mu$ we see from 
(\ref{eq:SNSymm-3}) that $b=-3$, $a=-5$, and hence $c=9/2$. 
To sum up, we have shown that 
\begin{equation}
\label{eq:SNSymm-4}
S_\mu\Psi(t,\vec x)=\mu^{9/2}\Psi(\mu^5t\,,\,\mu^3\vec x)
\end{equation}
satisfies the SN-equation for mass $m_\mu=\mu\cdot m$ 
if $\Psi$ satisfies the SN-equation for mass $m$.%
\footnote{Formulae (2.3) of \cite{Salzman-Carlip:2006}
and (2.4) of \cite{Carlip:2008} seem to be incorrect.}
\end{enumerate}

The stationary version of (\ref{eq:SchroedingerNewton}) is obtained by 
replacing $i\hbar\partial_t$ with the energy $E$. It has a unique 
(up to translations, of course) spherically symmetric stable ground 
state. The ground state minimises the energy functional which is 
given by the negative of the second and third term in 
(\ref{eq:NonLocalActionForSN}). 
Existence and uniqueness of the ground state were first shown 
by Lieb~\cite{Lieb:1977}, albeit motivated in completely different 
physical context (one-component plasmas) in 1976 by Choquard. 
Hence Lieb called  it the \emph{Choquard equation}, which is 
mathematically equivalent to the stationary SN-equation.

Numerical studies led to the following value for the 
ground-state energy\footnote{See \cite{RuffiniBonazzola:1969} 
for an early attempt and \cite{HarrisonMorozTod:2003} for 
a more recent and far more detailed account, also showing 
linear stability of the lowest and linear instability of all 
higher spherically symmetric stationary states found earlier 
in ~\cite{MorozPenroseTod:1998}. The 20 lowest energies 
are tabulated in Table\,1 of \cite{HarrisonMorozTod:2003} 
and range from $-0.163$ to $-0.000221$ times $G^2m^5/\hbar^2$.}
\begin{equation}
\label{eq:GroundStateEnergy}
E_0=-0.163\,\frac{G^2m^5}{\hbar^2}
=-0.163\cdot mc^2\cdot\left(\frac{m}{m_P}\right)^4\,,
\end{equation}   
where 
\begin{equation}
\label{eq:PlanckMass}
m_P:=\sqrt{\frac{\hbar c}{G}}
    =1.221\times 10^{19}\,\mathrm{GeV}/c^2
    =21.76\,\mathrm{\mu g}
\end{equation}
is the Planck mass. A rough intuition for the width, $a$, 
of the mass distribution $m\vert\Psi\vert^2$ in the ground
state is obtained by schematically writing the total energy 
as sum of the kinetic and potential term and replacing the 
Laplacian in the first by $a^{-2}$ and the double integral 
in the third term of (\ref{eq:NonLocalActionForSN}) by 
$a^{-1}$:
\begin{equation}
\label{eq:RoughEnergy}
E\approx\frac{\hbar^2}{2ma^2}-\frac{Gm^2}{2a}\,.
\end{equation}
Minimising in $a$ then gives
\begin{equation}
\label{eq:RoughGroundStateDiameter}
a_0\approx\frac{2\hbar^2}{Gm^3}
=2\ell_P\cdot\left(\frac{m_p}{m}\right)^3
\,,
\end{equation}
where 
\begin{equation}
\label{eq:PlanckLength}
\ell_P:=\sqrt{\frac{\hbar G}{c^3}}
    =1.616\times 10^{-35}\,\mathrm{m}
\end{equation}
is the Planck length. The rough energy (\ref{eq:RoughEnergy})
at the rough minimum (\ref{eq:RoughGroundStateDiameter}) is 
\begin{equation}
\label{eq:RoughGroundStateEnergy}
E_0\approx-\frac{1}{8}\cdot\frac{G^2m^5}{\hbar^2}
\end{equation}
which compares reasonably with (\ref{eq:GroundStateEnergy}).

Clearly, we would not trust the stationary solution in this 
Newtonian picture if the support of the mass distribution 
$m\vert\Psi\vert^2$ had a spatial extent not much larger than 
$m$'s Schwarzschild radius $Gm/c^2$. Now, the condition 
$a_0\gg Gm/c^2$ is equivalent to 
\begin{equation}
\label{eq:SanityCheck}
m^4\ll 2\cdot m_P^4\,, 
\end{equation}
which means that we should not let $m$ come too 
close to the Planck mass (though note the 4th powers).

Finally we note the dimensionless form of 
(\ref{eq:SchroedingerNewton}) that one obtains by 
introducing a spatial length scale $\ell$. 
Using dimensionless spatial coordinates ${\vec x}'$
and a dimensionless time $t'$ according to  
\begin{equation}
\label{eq:DimensionlessCoord}
{\vec x}':=\vec x/\ell\,,\qquad
t':=t\cdot\frac{\hbar}{2m\ell}
\end{equation}
and rescaling $\Psi\mapsto\Psi':=\ell^{3/2}\cdot\Psi$ so as to 
keep $\Psi'$ normalised with respect to $\D^3x'$, we get 
\begin{equation}
\label{eq:SchroedingerNewtonDimensionless}
i\partial_{t'}\Psi'(t',{\vec x}')=
\left(
-\Delta'-K\int\frac{\vert\Psi'(t',\vec y')\vert^2}%
{\Vert\vec x'-\vec y'\Vert}\,\D^3y'
\right)
\Psi'(t',\vec x')
\end{equation}
with dimensionless coupling constant 
\begin{equation}
\label{eq:DimensionlessCoupling}
K=2\cdot\frac{Gm^3 \ell}{\hbar^2}
=2\cdot\left(\frac{\ell}{\ell_P}\right)\left(\frac{m}{m_P}\right)^3\,.
\end{equation}
Note that the combination $(\text{length})\times(\text{mass})^3$
which determines the coupling had to be expected from scale 
invariance~(\ref{eq:SNSymm-4}).

\section{Analytical Estimations of the Collapse Mass}
We consider now the time-dependent SN-equation for initial 
values given by a spherically symmetric Gaussian wave packet
of width $a$:
\begin{equation}
\label{eq:GaussPacket}
\Psi(r,t=0) = (\pi a^2)^{-3/4} \, \exp \left(-\frac{r^2}{2 a^2}\right)\,.
\end{equation}
We have then two free parameters, $a$ and $m$,
and we ask for the regions in this two-parameters space in 
which significant inhibitions of the usual free dispersion 
occur. For simplicity we will refer to such a behaviour 
as a ``collapse'', even though it is clear that neither 
the integal over $\vert\psi\vert^2$ nor the volume of the 
support of  $\vert\psi\vert^2$ (which is infinite) change.
We start by presenting four different analytical 
arguments, all of which suggest that a collapse 
of the wave packet sets in whenever the dimensionless 
coupling $K$ in (\ref{eq:DimensionlessCoupling}) 
approaches unity.

\subsection{Equilibrium of Forces for the Free Solution}
Let us first give a rough estimate for the critical mass by a 
na\"{\i}ve calculation, in which we demand equilibrium between 
the acceleration of the peak probability density and the 
acceleration due to gravitation for a point mass.

The free solution (without gravitational potential) for the 
problem at hand is
\begin{equation}
\Psi_\text{free}(r,t) = \left(\pi a^2\right)^{-3/4}
\left(1 + \frac{i \,\hbar \, t}{m \, a^2}\right)^{-3/2} 
\, \exp \left( -\frac{r^2}{2 a^2 \, 
\left(1 + \frac{i \,\hbar \, t}{m \, a^2}\right)} \right).
\end{equation}
The radial probability density,  
$\rho(r,t)=4 \pi\,r^2 \,\abs{\Psi_\text{free}(r,t)}^2$,  
has a global maximum at
\begin{equation}
 r_p = a \,\sqrt{1 + \frac{\hbar^2 t^2}{m^2 a^4}}.
\end{equation}
Taking the second time derivative we obtain the peak's acceleration 

\begin{equation}
 \ddot{r}_p=\frac{\hbar^2}{m^2\,a^3} \, 
\left(1+\frac{\hbar^2\,t^2}{m^2\,a^4}\right)^{-3/2} 
= \frac{\hbar^2}{m^2\,r_p^3}\,.
\end{equation}
This we now compare to Newton's law for the gravitational 
acceleration 
\begin{equation}
 \ddot{r} = \frac{G \, m}{r^2}\,,
\end{equation}
at time $t=0$. Using $r_p(t = 0) = a$, we get
\begin{equation}
 m = \left( \frac{\hbar^2}{G \, a} \right)^{1/3}.
\end{equation}
For $a = \unit[0.5]{\mu m}$ this yields a threshold mass 
for collapse of about $\unit[7 \times 10^{-18}]{kg}$ or 
$\unit[4 \times 10^{9}]{u}$.

\subsection{Lower Mass Bound for Negative Energy Solutions}
It was noted by Harrison \emph{et al.}~\cite{HarrisonMorozTod:2003}
that only initial data of negative energy can show inhibitions of
dispersion. This can be used to provide a lower mass bound for
collapsing behaviour in a mathematically rigorous way. The result
strongly contradicts the collapse mass found in
\cite{Salzman-Carlip:2006,Carlip:2008}.

Define the kinetic energy $T$, the potential energy $V$,
and the total energy $\mathcal{E}$ respectively by
\begin{subequations}
\begin{align}
\mathcal{E} &= T + \frac{1}{2} \, V
    = \frac{\hbar^2}{2 m} \, \int \D^3 x\, \abs{\nabla \Psi(\vec x)}^2
     + \frac{m}{2}\, \int \D^3 x\, \Phi \, \abs{\Psi(\vec x)}^2 \nonumber\\
   &= \frac{\hbar^2}{2 m} \, \int \D^3 x\,
   \abs{\nabla \Psi(\vec x)}^2 - \frac{G m^2}{2} \int \D^3 x\,
   \int \D^3 y\, \frac{ \abs{\Psi(\vec x)}^2 \,
   \abs{\Psi(\vec y)}^2 }{\abs{\vec x - \vec y}}.
\end{align}
\end{subequations}
The second moment, defined by
\begin{equation}
Q = \int \abs{\vec{x}}^2 \, \abs{\Psi}^2 \, \D^3 x,
\end{equation}
then has the following second order time derivative:
\begin{equation}
\ddot{Q} = \frac{2 \hbar^2}{m^2} \, \int \abs{\nabla \Psi}^2 \, \D^3 x
  + \int \Phi \, \abs{\Psi}^2 \, \D^3 x
= \frac{1}{m} \, ( 4 T + V )
= \frac{1}{m} \, ( 4 \mathcal{E} - V ).
\end{equation}
But $\Phi$ (and therefore also $V$) is everywhere negative due to the
maximum principle which means that for positive energy $\mathcal{E}$,
$Q$ is a convex function of time. Hence, in terms of the second moment
it is a necessary (but not sufficient) condition for a collapsing
wave packet, that its initial data has negative energy.

We can now easily calculate the energy of the spherically symmetric,
real Gaussian wave packet \eqref{eq:GaussPacket}. It is
\begin{align}
\mathcal{E} &= \frac{2 \pi \,\hbar^2}{m} \,\int_0^\infty \D r \,
   (\partial_r \Psi(r))^2 - 8 \pi^2 \, G \, m^2 \nonumber\\
   &\bleq \times\int_0^\infty \D r \,\left[
   \int_0^r \D r' \, \frac{r'^2}{r} \,\Psi^2(r) \,\Psi^2(r')
   + \int_r^\infty \D r' \, r' \,\Psi^2(r) \,\Psi^2(r') \right]
   \nonumber\\
&= \frac{2 \hbar^2}{\sqrt{\pi} m \, a^7} \, \int_0^\infty \D r \,
   r^2 \e^{-r^2/a^2} - \frac{8 G m^2}{\pi\, a^6}
   \int_0^\infty \D r \, \frac{\e^{-r^2/a^2}}{r} \nonumber\\
   &\bleq \times \left[ \int_0^r \D r' \, r'^2 \,\e^{-r'^2/a^2}
   + r\, \int_r^\infty \D r' \, r' \,\e^{-r'^2/a^2} \right]
   \nonumber\\
&= \frac{\hbar^2}{2 m a^4} - \frac{2\,G\,m^2}{\sqrt{\pi} \,a^3}
   \,\sinh^{-1}(1).
\end{align}
Now $\mathcal{E} < 0$ is equivalent to
\begin{equation}
m > \left(\frac{\sqrt{\pi}\,\hbar^2}{4\,G\,a\,\sinh^{-1}(1)}
    \right)^{1/3}
\approx \left(\frac{\hbar^2}{2 \,G\,a}\right)^{1/3}
\end{equation}
By inserting $a = \unit[0.5]{\mu m}$ again this yields the lower
mass bound for negative energy solutions
$m > \unit[5.5 \times 10^{-18}]{kg} \approx
\unit[3.3 \times 10^{9}]{u}$.

\subsection{Mass of the Stationary Ground State}
In our third example we consider the ground state of the 
spherically symmetric SN-equation found by Moroz \emph{et al.} 
in~\cite{MorozPenroseTod:1998}. In order to do this we 
need to rescale the SN-equation in a suitable fashion in order to 
to compare our form of the equation, which explicitly contains 
the mass parameter, with the form solved 
in~\cite{MorozPenroseTod:1998}, in which $m$ does not appear.
In order to understand how to transform solutions of one form 
into solutions of the other, we use scale invarianvce. 

In a first step we substitute $U=\frac{2 m}{\hbar^2} (E-m \Phi)$ 
into the stationary SN-equation
\begin{subequations}
\label{eq:stationary_SN}
\begin{align}
E \Psi(\vec{x}) 
&= \left[-\frac{\hbar^2}{2 m} \Delta + m \Phi(\vec{x})\right] \,
\Psi(\vec{x})\,,\\
\Delta \Phi(\vec{x}) &= 4 \pi \, G \, m \, \abs{\Psi(\vec{x})}^2\,.
\end{align}
\end{subequations}
It then takes the following form:
\begin{subequations}
\label{eq:StationarySN}
\begin{align}
U(\vec{x}) \Psi(\vec{x}) &= -\Delta \Psi(\vec{x})\,, 
\label{eq:U_Psi_Delta_Psi}\\
\Delta U(\vec{x}) &= -\frac{2 m^2}{\hbar^2} \Delta \Phi(\vec{x})
= -\frac{8 \pi \, G \, m^3}{\hbar^2} \, \abs{\Psi(\vec{x})}^2\,.
\end{align}
\end{subequations}
This can be further simplyfied if written in terms of the 
rescaled function $\Omega = \alpha \Psi$, 
where $\alpha:=\sqrt{8\pi Gm^3}/\hbar$. We get
\begin{subequations}
\label{eq:SN-equation_for_Omega}
\begin{align}
\Delta \Omega(\vec{x}) &= -U(\vec{x})\, \Omega(\vec{x})\,,\\
\Delta U(\vec{x}) &= -\abs{\Omega(\vec{x})}^2\,.
\end{align}
\end{subequations}
The norm of $\Omega$ is obviously
$\Vert\Omega\Vert = \alpha \,\Vert\Psi\Vert$.

Now we rescale the mass by a factor of $\mu$, as discussed above. 
Then, according to equation~\eqref{eq:SNSymm-4}, 
$S_\mu\Psi(\vec x)=\mu^{9/2}\Psi(\mu^3\vec x)$ 
and $\alpha \to \alpha_\mu = \mu^{3/2} \alpha$. Therefore
\begin{equation}\label{eq:S_mu_Omega}
\begin{array}{rclclcl}
S_\mu\Omega(\vec x) &=& \alpha_\mu \, S_\mu \Psi(\vec x) &=& \mu^6 \alpha \Psi(\mu^3\vec x)
&=& \mu^6 \Omega(\mu^3\vec x),\\
\Vert S_\mu\Omega\Vert &=& \alpha_\mu \, \Vert S_\mu \Psi \Vert
&=& \mu^{3/2} \,\alpha\, \Vert\Psi\Vert &=& \mu^{3/2} \, \Vert\Omega\Vert.
\end{array}
\end{equation}
This means that the rescaling changes the norm of $\Omega$ and 
the following statement holds:
\begin{proposition}
Suppose that $(\Omega,\,U)$ satisfy \eqref{eq:SN-equation_for_Omega}. For a given mass
value $m$ let $\mu$ be the rescaling parameter for which the norm of $S_\mu \Omega$ is
\begin{equation}
\Vert S_\mu \Omega\Vert = \frac{\sqrt{8 \pi G \, m^3}}{\hbar}.
\end{equation}
Then the wave function defined by
\begin{equation}\label{eq:Normalised_Wavefunction_from_Omega}
\Psi(\vec x) := \frac{\hbar}{\sqrt{8 \pi G \, m^3}} \, S_\mu \Omega(\vec x)
\end{equation}
is a normalised solution to the stationary
SN-equation~\eqref{eq:stationary_SN} with mass parameter $m$.
\end{proposition}
Putting equations~\eqref{eq:S_mu_Omega}--\eqref{eq:Normalised_Wavefunction_from_Omega} 
together we obtain a formula that gives us a normalised 
wave function satisfying the stationary SN-equation (\ref{eq:StationarySN})
with mass parameter $m$ for any unnormalised solution $\Omega$ of 
(\ref{eq:SN-equation_for_Omega}):
\begin{equation}
\Psi(\vec x) = \frac{(8 \pi G)^{3/2}\,m^{9/2}}{\hbar^3 \, \Vert\Omega\Vert^4}\,
\Omega\left( \frac{8 \pi G m^3}{\hbar^2 \,\Vert\Omega\Vert^2}\,\vec x \right).
\end{equation}

\begin{figure}[t]
\centering
\includegraphics[width=12cm]{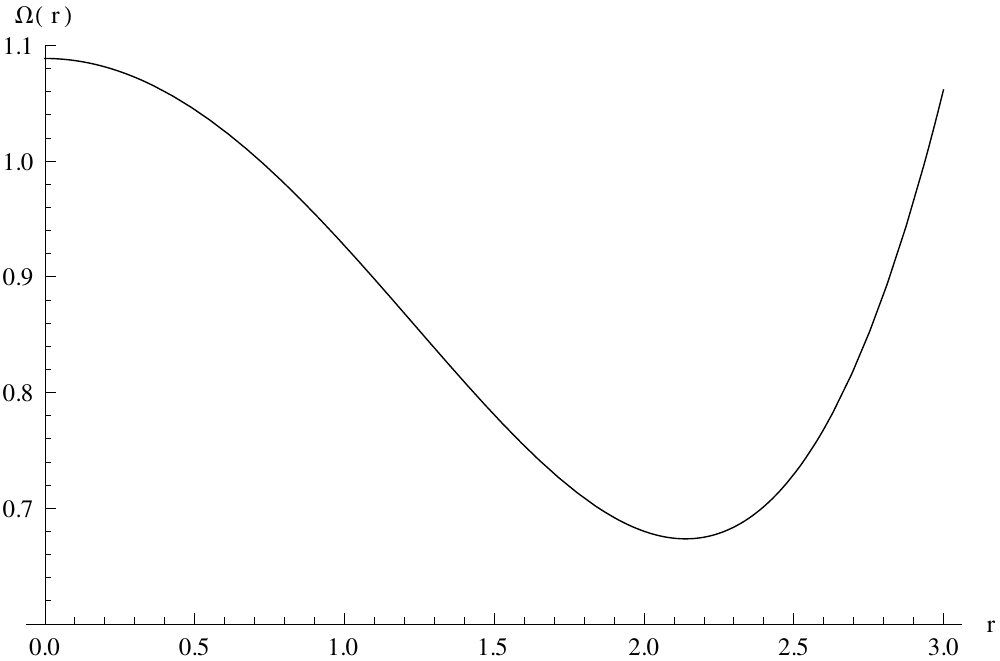}
\caption{The ground state solution \eqref{eq:GroundStateSolutionMoroz} with
$\Omega_0 = 1.08864$ found by Moroz \emph{et al.}~\cite{MorozPenroseTod:1998}
plotted up to forth order. It seems reasonable from this plot to cut off
the solution for values of $r > 2.2$.}
\label{fig:omega}
\end{figure}

\noindent In \cite{MorozPenroseTod:1998} Moroz \emph{et\,al.}
found spherically symmetric, real solutions\footnote{For the spherically symmetric
SN-equation, the solution can w.l.o.g. be assumed real. This follows from \eqref{eq:U_Psi_Delta_Psi} if restricted to the spherically symmetric case.}
\begin{equation} \label{eq:GroundStateSolutionMoroz}
\Omega(r) = \Omega_0 \, \left( 1 - \frac{1}{6}\, r^2 + \frac{\Omega_0^2 + 1}{120}\, r^4
+ \dots \right)
\end{equation}
to (\ref{eq:StationarySN}), where $\Omega_0 = \Omega(0)$ and $\Omega_0 \approx 1.08864$ 
for the ground state. These polynomial functions are in general not normalisable but 
for the right choice of $\Omega_0$ they  tend to zero asymptotically as $r \to \infty$.
Nevertheless, for the approximate solution we must introduce a cutoff $\Lambda$ such that
$\Omega(r) \equiv 0$ for $ r > \Lambda$. The ground state solution is plotted in
figure~\ref{fig:omega} up to order $r^4$, according to which $\Lambda \approx 2.2$ 
seems to be a reasonable choice. Therefore
\begin{equation}
\Vert\Omega\Vert^2 \approx \int_0^\Lambda 4 \pi r^2 \D r \, \Omega^2(r) \approx 26.
\end{equation}
If we define as the width of $\Omega$ the cutoff $\Lambda$, the width of
the wave packet $\Psi$ is
\begin{equation}
a = \frac{\hbar^2 \,\Vert\Omega\Vert^2}{8 \pi G m^3} \, \Lambda
\approx \frac{57 \, \hbar^2}{8 \pi G m^3},
\end{equation}
or written in terms of $m$
\begin{equation}
m \approx \left(\frac{57 \, \hbar^2}{8 \pi G a}\right)^{1/3}.
\end{equation}
With the usual value of $a = \unit[0.5]{\mu m}$ we get a mass of
\begin{equation}
 m \approx \unit[9.1 \times 10^{-18}]{kg} \approx \unit[5.5 \times 10^{9}]{u}.
\end{equation}

\subsection{Short-Time Behaviour}
Finally, to obtain the behaviour of the spherically symmetric
SN-equation with an initial
Gaussian wave packet for small times, we consider the lowest
order terms of the series expansion
around $t=0$.
Transforming to $\phi(r,t) := r \, \Psi(r,t)$ and considering
the spherically symmetric gravitational potential
\begin{equation}
\Phi(r,t) = -4 \pi \, G \, m \, \left[ \frac{1}{r} \int_0^r 
\abs{\phi(r',t)}^2 \D r' 
+ \int_r^\infty \frac{\abs{\phi(r',t)}^2}{r'} \D r' \right],
\end{equation}
the SN-equation together with its complex conjugate reads
\begin{equation}
\partial_t \phi = \frac{i \hbar}{2 m} \, \partial_r^2 \phi 
- \frac{i\,m}{\hbar} \, \Phi \, \phi, \quad\quad
\partial_t \phi^* = -\frac{i \hbar}{2 m} \, \partial_r^2 \phi^* 
+ \frac{i\,m}{\hbar} \, \Phi \, \phi^*,
\end{equation}
where, as before, the star denotes complex conjugation.
We get the first order time derivative
\begin{equation}
 \partial_t \abs{\phi}^2 = (\partial_t \phi) \, \phi^* 
+ \phi \, (\partial_t \phi^*) = \frac{i \hbar}{2 m}
\left((\partial_r^2 \phi) \, \phi^* - \phi \, (\partial_r^2 \phi^*)\right)
\end{equation}
of the absolute value squared which vanishes at $t=0$ because $\phi(r,0)$ is real valued.
For the second order time derivatives we obtain
\begin{subequations}
\begin{align}
 \partial_t^2 \phi &= \frac{i \hbar}{2 m} \, \partial_r^2 (\partial_t \phi) 
- \frac{i\,m}{\hbar} \, (\partial_t \Phi) \, \phi 
- \frac{i\,m}{\hbar} \, \Phi \, (\partial_t \phi) \nonumber\\
&= -\frac{\hbar^2}{4 m^2} \, \partial_r^4 \phi 
+ \frac{1}{2}\, (\partial_r^2 \Phi) \phi 
+ (\partial_r \Phi) (\partial_r \phi) 
+ \Phi \,(\partial_r^2 \phi) \nonumber\\
&\bleq - \frac{m^2}{\hbar^2} \Phi^2 \phi - \frac{i\,m}{\hbar} (\partial_t \Phi) \phi \\
\partial_t^2 \phi^* &= -\frac{\hbar^2}{4 m^2} \, \partial_r^4 \phi^*
+ \frac{1}{2}\, (\partial_r^2 \Phi) \phi^* 
+ (\partial_r \Phi) (\partial_r \phi^*) 
+ \Phi \,(\partial_r^2 \phi^*) \nonumber\\
&\bleq - \frac{m^2}{\hbar^2} \Phi^2 \phi^* + \frac{i\,m}{\hbar} (\partial_t \Phi) \phi^* \\
\partial_t^2 \abs{\phi}^2 &= (\partial_t^2 \phi)\, \phi^* 
+ 2\, (\partial_t \phi) \,(\partial_t \phi^*) 
+ \phi\, (\partial_t^2 \phi^*) \nonumber\\
&= -\frac{\hbar^2}{4 m^2} \left[ (\partial_r^4 \phi) \, \phi^* 
- 2 \,(\partial_r^2 \phi) \, (\partial_r^2 \phi^*) 
+ \phi \, (\partial_r^4 \phi^*) \right]
\nonumber\\ 
&\bleq + (\partial_r^2 \Phi) \, \abs{\phi}^2 
+ (\partial_r \Phi) \, \left( (\partial_r \phi) \, \phi^* 
+ \phi \, (\partial_r \phi^*) \right). \label{eqn:secord}
\end{align}
\end{subequations}
We introduce some shorthand notations
\begin{subequations}
\begin{align}
\alpha &:= (\pi a^2)^{-3/2} & \beta &:= a^{-1} & \rho &:= \beta r \\
\gamma_1 &:= 4 \pi^{-1/2} \, G \, \alpha \, \beta &
\gamma_2 &:= 2 \, G \, \alpha \, \beta &
\gamma_3 &:= \hbar^2 \, \alpha \, \beta^2 \\
\widetilde{\Phi} &:= -\alpha \, \gamma_1^{-1} \, m^{-1} \, \Phi &
\phi_0(r) &:= \phi(r,0) = \sqrt{\alpha} \, r \, \e^{-\beta^2 r^2 / 2}, \hspace{-\textwidth}
\end{align}
\end{subequations}
and consider the value of $\partial_t^2 \abs{\phi}^2$ at $t=0$ which,
as $\Phi \sim m$, takes the form 
\begin{align}
 \left. \partial_t^2 \abs{\phi}^2 \right|_{t=0} &= m \, A(\rho) + \frac{1}{m^2} \, B(\rho), \\
\intertext{where}
 A(\rho) &= -\frac{\gamma_1}{\alpha} \left((\partial_r^2 \widetilde{\Phi}) \, \phi_0^2 + 2 \,
 (\partial_r \widetilde{\Phi}) \, (\partial_r \phi_0) \, \phi_0 \right) \label{eqn:Aofrho}\\
 B(\rho) &= -\frac{\hbar^2}{2} \left( \phi_0 \, (\partial_r^4 \phi_0)
- (\partial_r^2 \phi_0)^2 \right).
\end{align}
Evaluating the derivatives of $\widetilde{\Phi}$ we obtain
\begin{subequations}
\begin{align}
 \widetilde{\Phi} &= \frac{1}{\rho} \, \int_0^\rho \rho'^2 \, \e^{-\rho'^2} \D\rho' + 
 \int_\rho^\infty \rho' \, \e^{-\rho'^2} \D\rho' \nonumber\\
 &= \frac{\sqrt{\pi}}{4}\, \frac{\mathrm{erf}(\rho)}{\rho} \\
\frac{1}{\beta}\,\partial_r \widetilde{\Phi} = \partial_\rho \widetilde{\Phi}
&= \frac{1}{2 \rho} \, \e^{-\rho^2} - \frac{\sqrt{\pi}}{4}\,
\frac{\mathrm{erf}(\rho)}{\rho^2} \label{eqn:drV} \\
\frac{1}{\beta^2}\,\partial_r^2 \widetilde{\Phi} = \partial_\rho^2 \widetilde{\Phi}
&= \frac{\sqrt{\pi}}{2}\, \frac{\mathrm{erf}(\rho)}{\rho^3}
- \,\left( 1 + \frac{1}{\rho^2} \right) \, \e^{-\rho^2},\label{eqn:drV2}
\end{align}
\end{subequations}
where $\mathrm{erf}(\rho)$ is the \emph{error function}
\begin{equation}
 \mathrm{erf}(\rho) := \frac{2}{\sqrt{\pi}} \, \int_0^\rho \e^{-x^2} \D x.
\end{equation}

% Somewhat misplaced, but the figures shoud be in this subsection, thus we put them here
\newcommand{\pltw}{150pt}
\newcommand{\pltsp}{\quad\quad\quad}
\begin{figure}[t]
 \centering
 \subfloat[$A(\rho)$]{\includegraphics[width=\pltw]{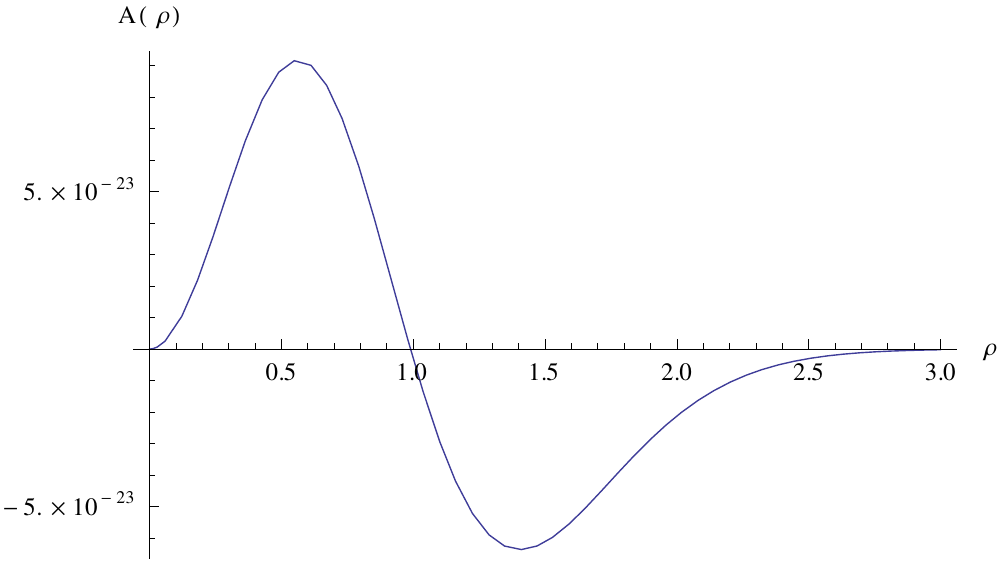}} \pltsp
 \subfloat[$B(\rho)$]{\includegraphics[width=\pltw]{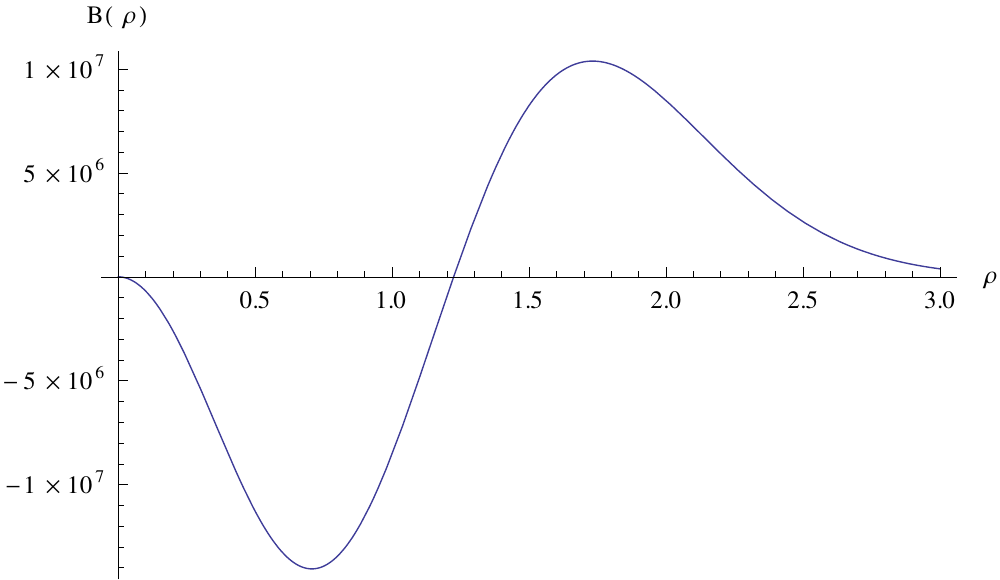}}
 \caption{Components $A(\rho)$ and $B(\rho)$ of the order $t^2$ contribution
 to $\abs{\phi}^2$ which is proportional to
 $m \, A(\rho)  + \frac{1}{m^2} \, B(\rho)$.
 While $A$ is attractive, $B$ shows a repulsive behaviour.}
 \label{fig:plot_AB}
\end{figure}

\begin{figure}[p]
 \centering
 \subfloat[$m = 10^6 \unit{u}$]{
  \includegraphics[width=\pltw]{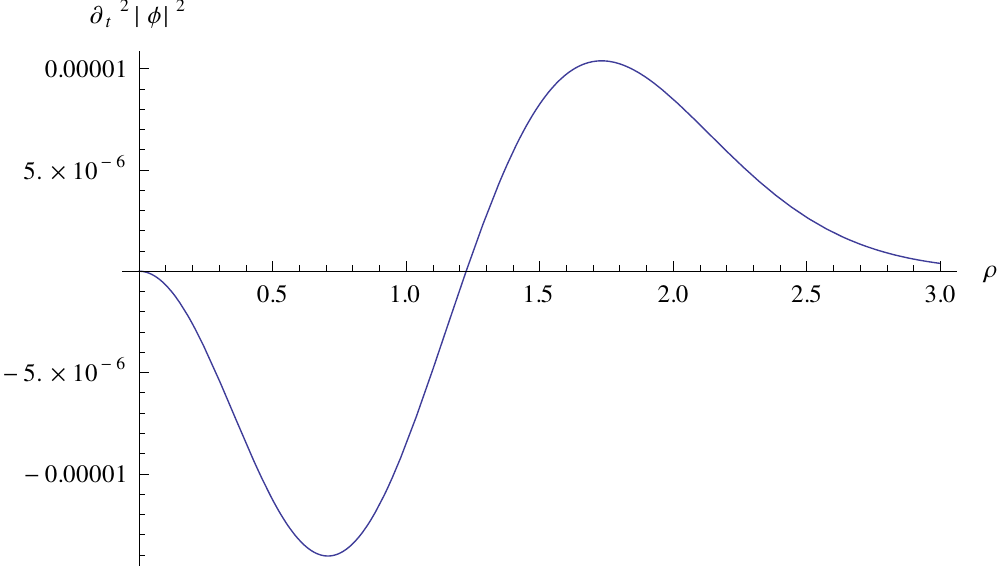}}\pltsp
 \subfloat[$m = 4 \times 10^9 \unit{u}$]{
  \includegraphics[width=\pltw]{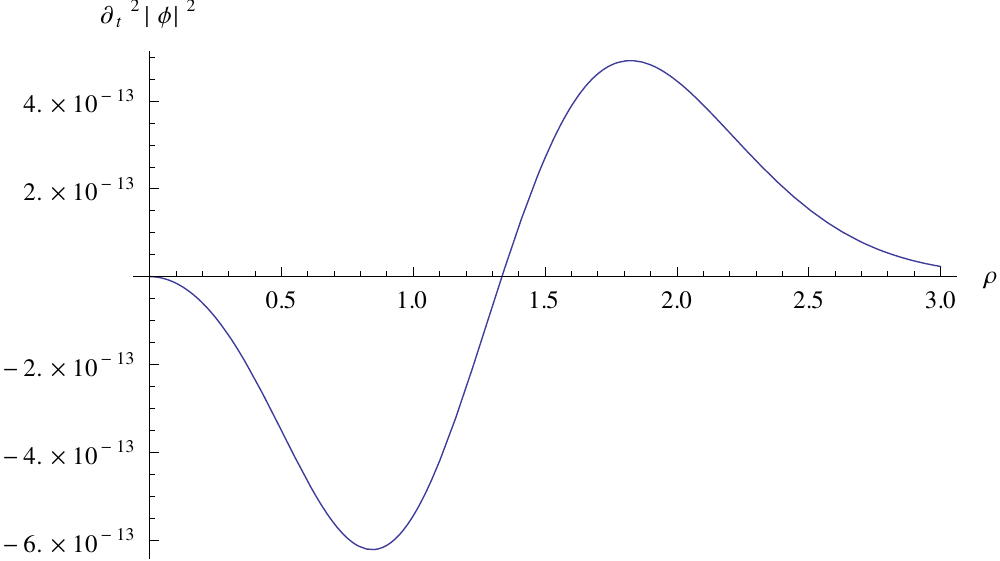}}\\
 \subfloat[$m = 5 \times 10^9 \unit{u}$]{
  \includegraphics[width=\pltw]{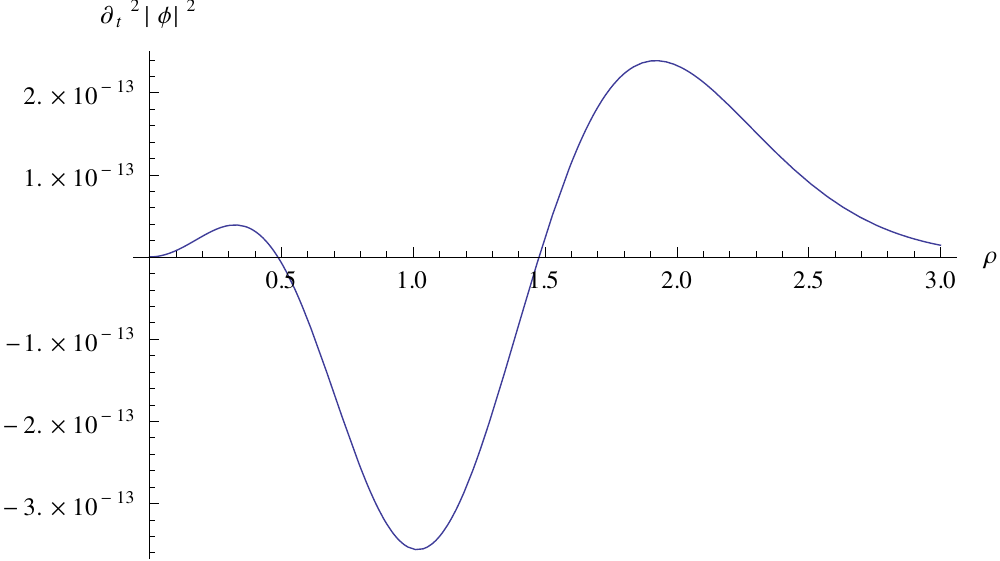}}\pltsp
 \subfloat[$m = 6 \times 10^9 \unit{u}$]{
  \includegraphics[width=\pltw]{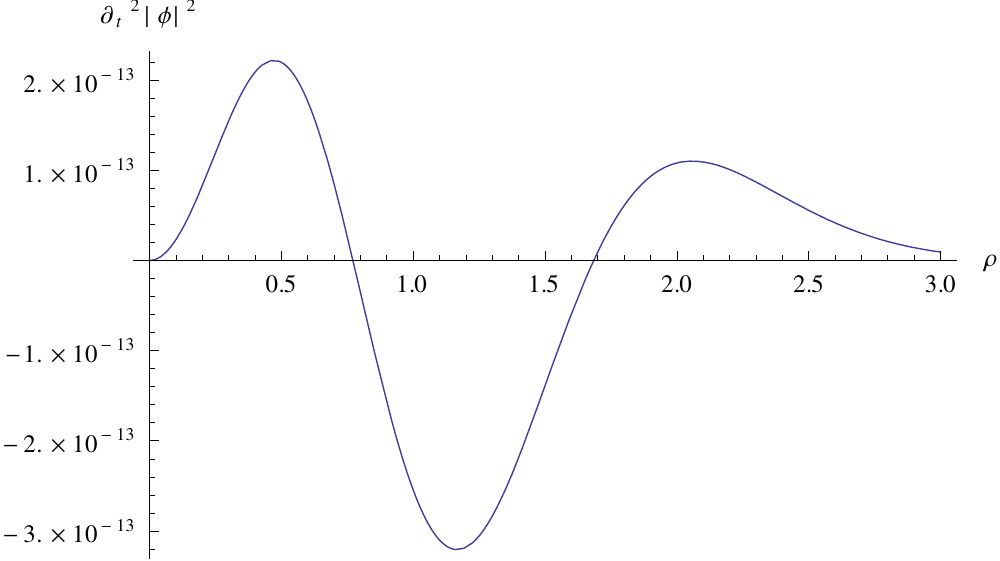}}\\
 \subfloat[$m = 7 \times 10^9 \unit{u}$]{
  \includegraphics[width=\pltw]{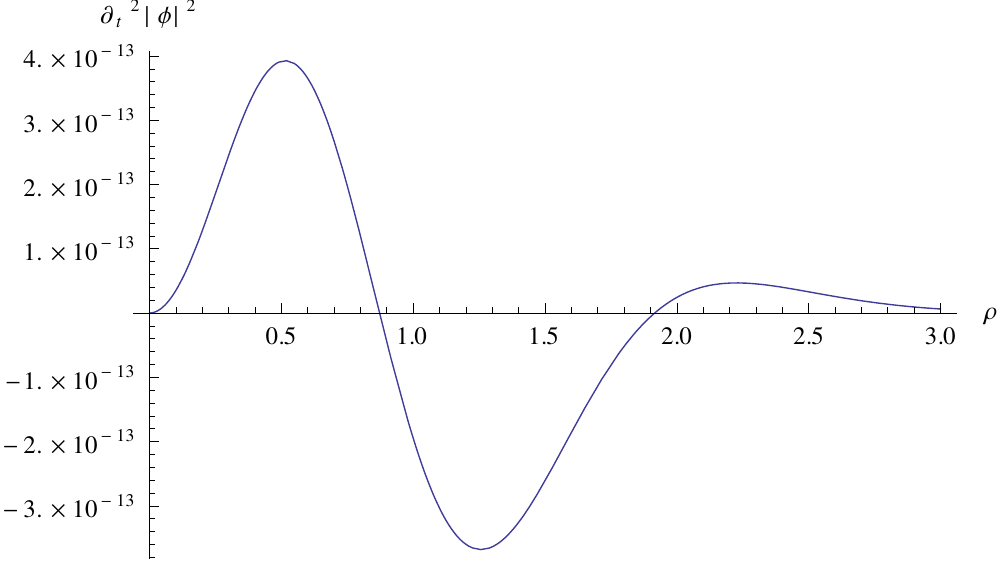}}\pltsp
 \subfloat[$m = 8 \times 10^9 \unit{u}$]{
  \includegraphics[width=\pltw]{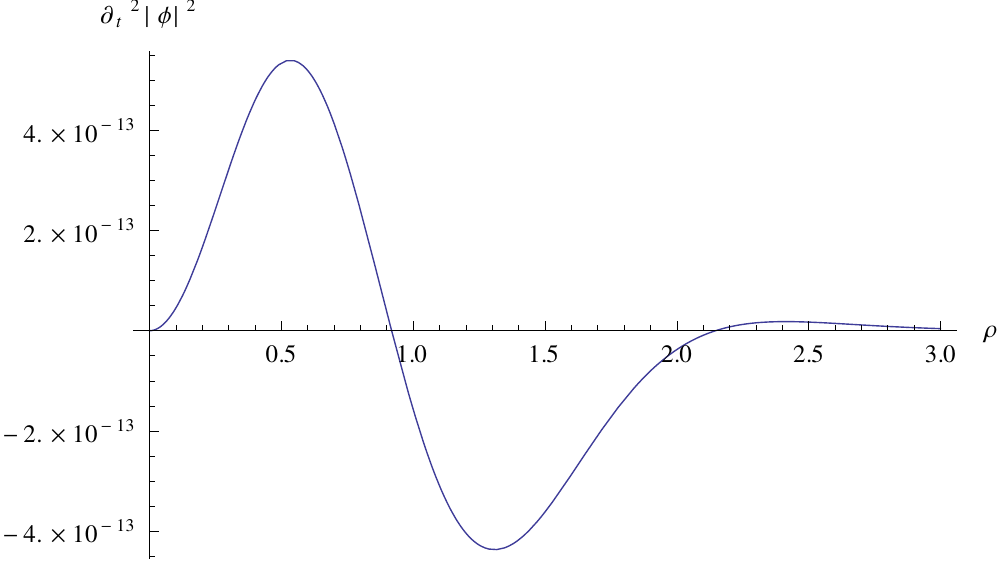}}\\
 \subfloat[$m = 9 \times 10^9 \unit{u}$]{
  \includegraphics[width=\pltw]{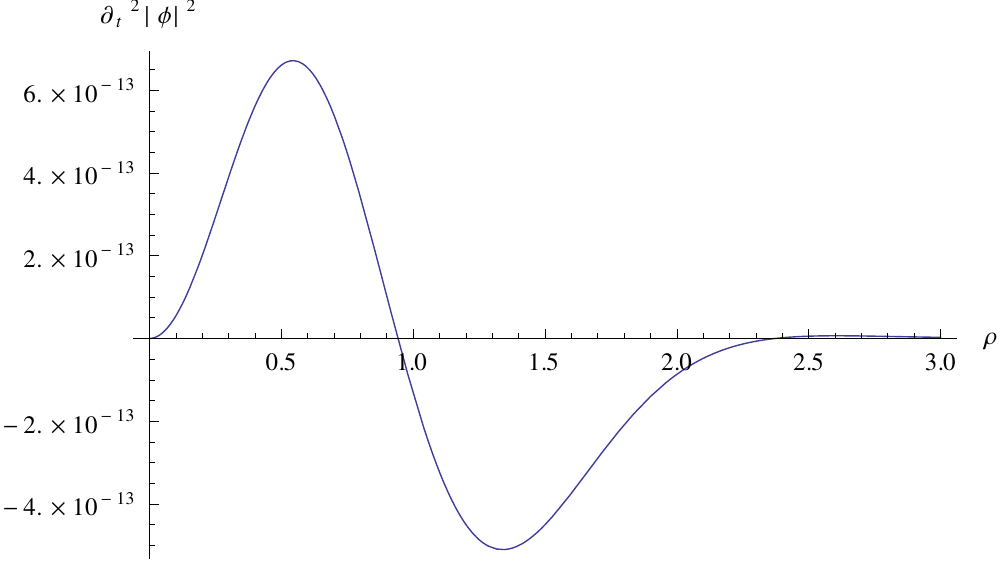}}\pltsp
 \subfloat[$m = 10^{12} \unit{u}$]{
  \includegraphics[width=\pltw]{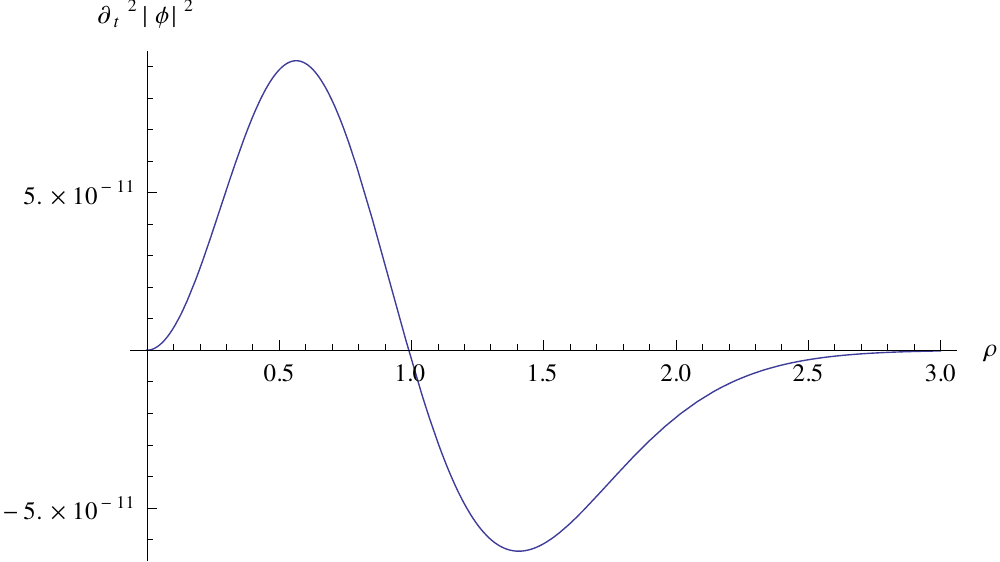}}
 \caption{Plot of $\partial_t^2 \abs{\phi}^2$ at $t=0$ for several masses.
  While the graph changes only in its quantitative scale and
  does not change qualtitatively in its shape for masses smaller than
  $m = 4 \times 10^9 \unit{u}$ and larger than
  $m = 9 \times 10^9 \unit{u}$, respectively, one can see a
  qualitative change from repulsive to attractive behaviour
  in between those mass values.}
 \label{fig:plot_phi}
\end{figure}

% end of figures, continue with text

\noindent Inserting \eqref{eqn:drV} and \eqref{eqn:drV2} into~\eqref{eqn:Aofrho},
the final result can now be stated as
\begin{align}
 \abs{\phi}^2 &= \phi_0^2 + \frac{t^2}{2}\, \left( m \, A(\rho) 
+ \frac{1}{m^2} \, B(\rho) \right) + \mathcal{O}(t^3), \\
\intertext{where}
 A(\rho) &= 2 \, \gamma_1 \, \rho^2 \, \e^{-2 \rho^2} 
- \gamma_2 \, \rho \, \e^{-\rho^2} \, \mathrm{erf}(\rho) \\
 B(\rho) &= \gamma_3 \, \rho^2 \, (2 \rho^2 - 3) \, \e^{-\rho^2}.
\end{align}
$A(\rho)$ and $B(\rho)$ are plotted in figure \ref{fig:plot_AB}.
While $A$ shows an attractive behaviour (i.\,e. the wave function
grows for small values of $\rho$), $B$ shows a repulsive behaviour.
Thus, the attractive behaviour gets dominant for large masses, as expected.

In figure \ref{fig:plot_phi} the second order derivative of $\abs{\phi}^2$ 
is plotted for several values of the mass $m$. One can see that there is a 
repulsive behaviour for $m \lesssim 4 \times 10^9 \,\unit{u}$ which turns 
with increasing mass and is completely attractive for masses 
$m \gtrsim 9 \times 10^9 \,\unit{u}$. We therefore expect the collapse of 
the wave function to happen at about 5--8 $\times 10^9 \,\unit{u}$.

\section{Numerical Investigation of the time dependent 
Schr\"o\-din\-ger-New\-ton Equation}

Our numerical methods do not differ in principle from those used in
\cite{Salzman-Carlip:2006}. A Crank-Nicolson scheme is applied to solve
the differential equation \eqref{eq:SchroedingerNewton} in a spherically
symmetric context.

Having defined both a spatial and temporal grid size $\Delta r$ and
$\Delta t$ and using the index notation
$\Psi^n_j = \Psi(j \Delta r, n \Delta t)$ we can write equation
\eqref{eq:SchroedingerNewton} in a discretised way using Cayley's form
of the SN-equation
\begin{equation}
\exp \left(\frac{i \, \Delta t}{2 \hbar} H\right) \, \Psi^{n+1}_j
= \exp \left(-\frac{i \, \Delta t}{2 \hbar} H\right) \, \Psi^{n}_j.
\end{equation}
Linearising this equation we can write it as
%
%%%%%%% Figures hier damit sie an der richtigen Stelle erscheinen %%%
%
\begin{figure}[t]
\centering
\includegraphics[width=12cm]{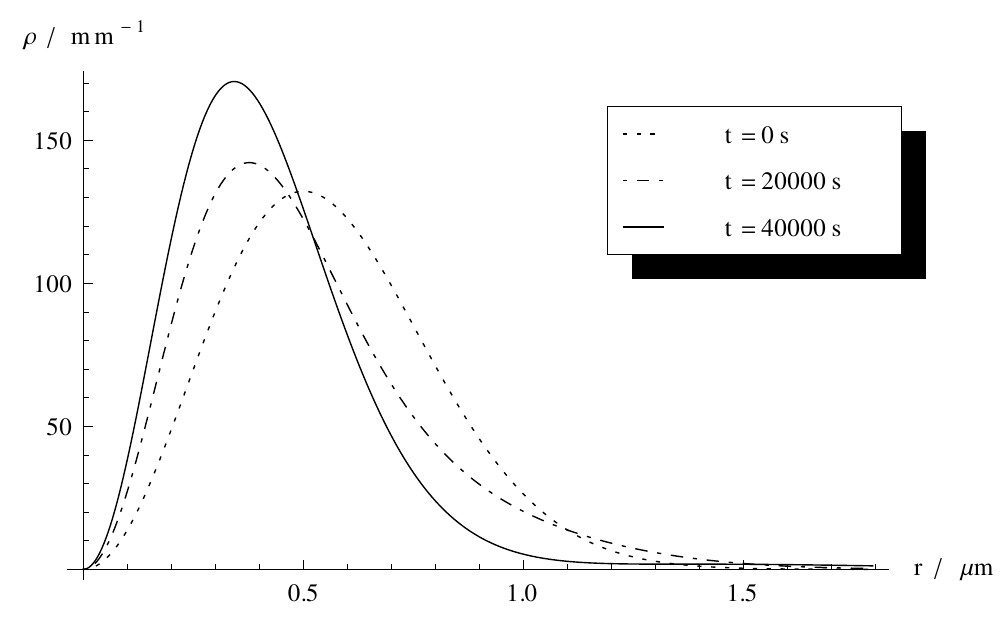}
\caption{Collapsing wave packet for $m = 7 \times 10^9\,\mathrm{u}$.
  Plotted is the radial probability density $\rho = 4 \pi \, r^2 \, \abs{\psi}^2$
  against $r$ at three different times.}
\label{fig:collapse}
\end{figure}
\begin{figure}[t]
\centering
\includegraphics[width=12cm]{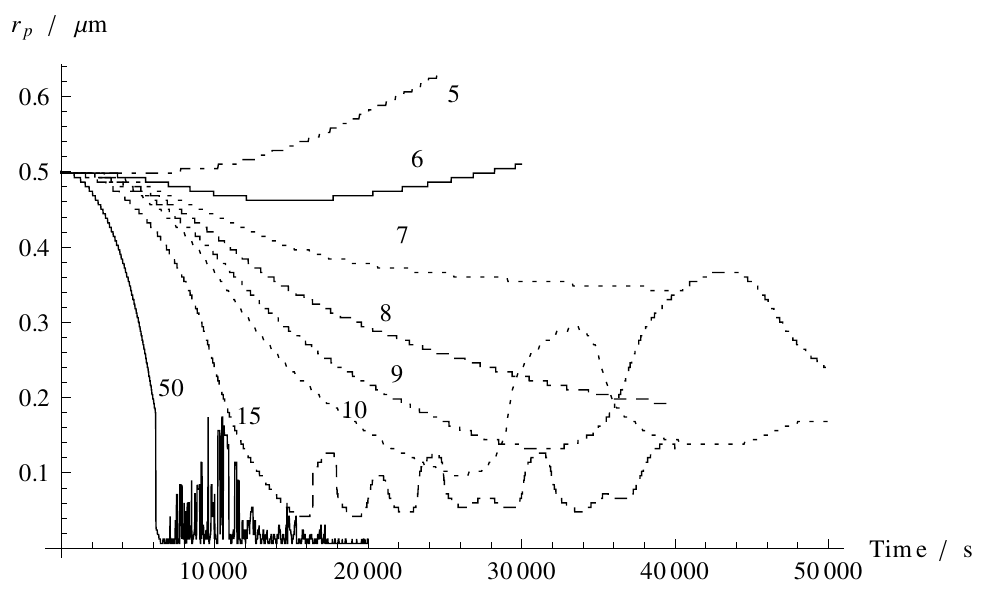}
\caption{The peak of the radial probability density plotted against time for
several masses. The curves are marked with the mass in units of
$10^9\,\mathrm{u}$.}
\label{fig:rp}
\end{figure}
%
%%%%%%%%%%%%%%%%%%%%%%%%%%%%%%%%%%%%%%%%%%%%%%%%%%%%%%%%%%%%%%%%%%%%%
%
\begin{subequations}
\begin{align}
\Psi^{n+1}_j &= (Q^{-1} - \mathbbm{1}) \Psi^{n}_j \\
\label{eqn:Q matrix}
Q &= \frac{1}{2} \, \left(\mathbbm{1}
     + \frac{i \, \Delta t}{2 \hbar} \, H\right).
\end{align}
\end{subequations}
We thus have to solve the linear system
\begin{align}
\label{eqn:linear system}
Q \, \chi^n &= \Psi^n\\
\intertext{to obtain}
\Psi^{n+1} &= \chi^n - \Psi^n.
\end{align}
The radial component of the Laplacian in spherical coordinates,
\begin{equation}
\Delta_r = \left\{\begin{array}{ll} \frac{\partial^2}{\partial r^2}
           + \frac{2}{r} \frac{\partial}{\partial r}
           & \quad \text{if } r > 0\\ 3 \frac{\partial^2}{\partial r^2}
           &  \quad \text{if } r = 0, \end{array}\right.
\end{equation}
takes the discretised form
\begin{equation}
\label{eqn:discrete laplacian}
\Delta \chi^n_j = \left\{\begin{array}{ll} \frac{1}{(\Delta r)^2} \,
                  \left( \frac{j-1}{j}\, \chi^n_{j-1} - 2 \chi^n_{j}
                  + \frac{j+1}{j}\, \chi^n_{j+1} \right)
                  & \quad \text{if } j > 0\\ \frac{1}{(\Delta r)^2} \,
                  \left( -6 \chi^n_0 + 6 \chi^n_1 \right)
                  &  \quad \text{if } j = 0, \end{array}\right.
\end{equation}
while the discretised form of the potential
\begin{subequations}
\begin{align}
\Phi &= -4 \pi G m \left( \frac{1}{r} \int_0^r \abs{\Psi(r',t)}^2 r'^2 \, \D r'
     + \int_r^\infty \abs{\Psi(r',t)}^2 r' \, \D r' \right)
\intertext{is}
\label{eqn:discrete potential}
\Phi^n_j &= -4 \pi G m (\Delta r)^2 \, v^n_j \\
v^n_j &= \frac{1}{j} \sum_{i=0}^{j-1} \abs{\Psi^n_i}^2 \, i^2
         + \sum_{i=j}^{N-1} \abs{\Psi^n_i}^2 \, i.
\end{align}
\end{subequations}
$Q$ then becomes a tridiagonal matrix
\begin{equation}
Q = \begin{pmatrix}
b_0 & c_0 & 0   & 0 	  & \cdots \\
a_1 & b_1 & c_1 & 0       & \cdots \\
0   & a_2 & b_2 & c_2     &  \\
\vdots &  &     & \ddots  & \vdots \\ 
0   & \cdots &  & a_{N-1} & b_{N-1}
\end{pmatrix},
\end{equation}
where
\begin{subequations}
\begin{align}
&& a_j &= \beta \, \frac{j-1}{j} & (0 < j \leq N-1) \\ 
   b_0 &= \frac{1}{2} - 6 \beta - \gamma\, v_0
&  b_j &= \frac{1}{2} - 2 \beta - \gamma\, v_j  & (0 < j \leq N-1) \\
   c_0 &= 6 \beta & c_j &= \beta \, \frac{j+1}{j}  & (0 < j < N-1)
\end{align}
\end{subequations}
using the shorthand notations
\begin{equation}
\beta = -\frac{i \hbar}{8 m} \, \frac{\Delta t}{(\Delta r)^2}
\quad \quad \text{and} \quad \quad
\gamma = \frac{i \pi G}{\hbar} \, m^2 \, \Delta t \, (\Delta r)^2.
\end{equation}
We use the tridiagonal matrix algorithm to solve the linear system
\eqref{eqn:linear system}.

\subsection{Numerical Results}
We start with a spherically symmetric Gaussian wave packet 
(\ref{eq:GaussPacket}) whose width is always set to a value 
of $a = 0.5\,\mu\mathrm{m}$. Note that due to the scaling 
law \eqref{eq:SNSymm-4} the width can be fixed without 
loss of generality.

\begin{figure}[t]
\centering
\includegraphics[width=12cm]{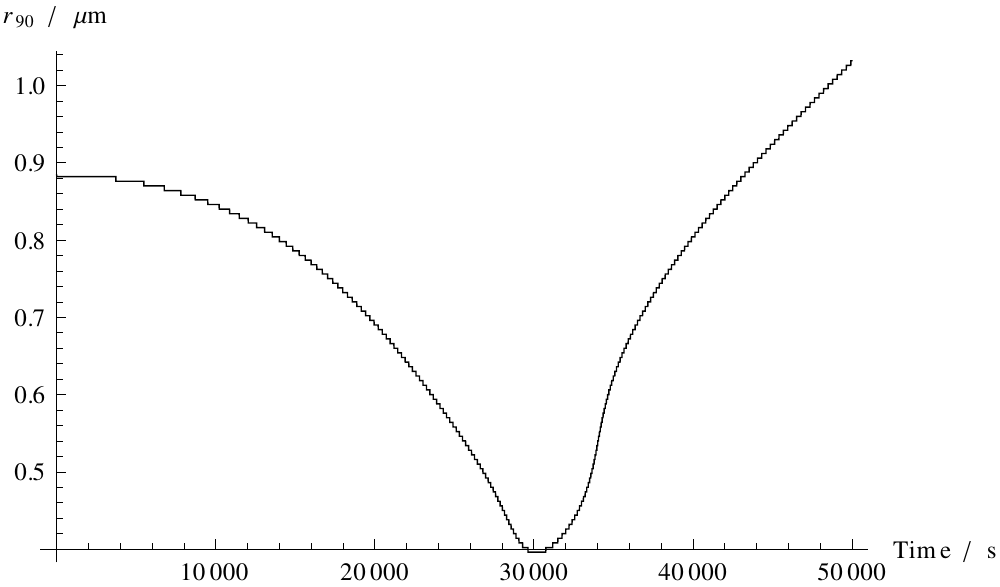}
\caption{The radius $r_{90}$ within which 90\,\% of the
probability lie plotted against time for a mass of
$m=10^{10}\,\mathrm{u}$. Note that the minimum is
\emph{not} at zero but at about $r_{90}\approx0.4\;\mathrm{\mu m}$.}
\label{fig:r90plot}
\end{figure}

\begin{figure}[p]
\centering
\subfloat[full mass range]{\includegraphics[width=12cm]{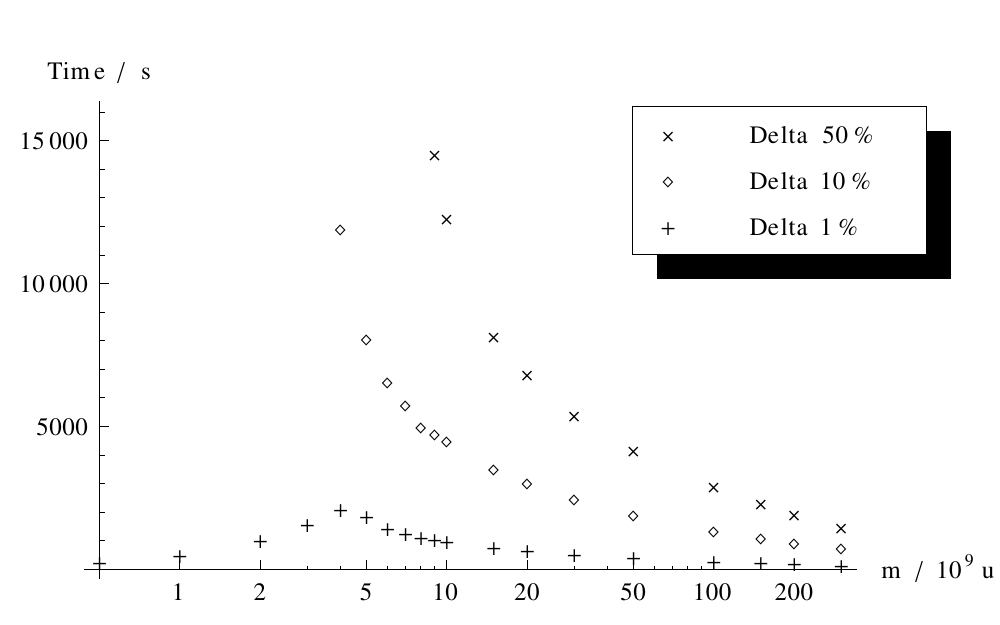}}\\
\subfloat[largest masses]{\includegraphics[width=12cm]{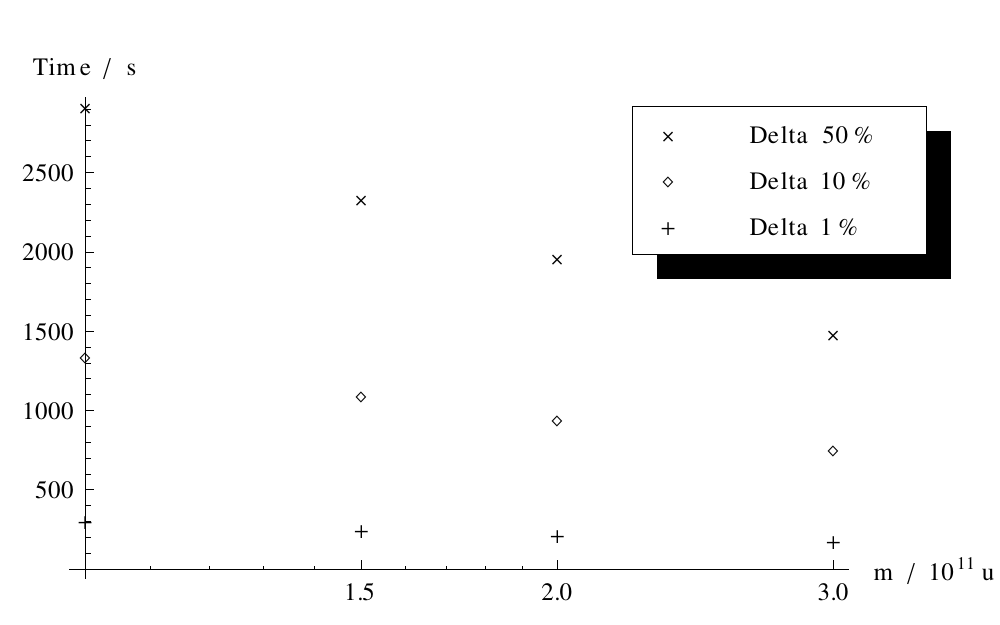}}
\caption{Time it takes until the gravitationally interacting solution differs from the
solution of the free Schr\"odinger equation in its FWHM by a percentage of 1\,\%, 10\,\%,
and 50\,\% respectively.}
\label{fig:mass_time}
\end{figure}

The results can be summarised as follows:
\begin{itemize}
\item For masses $m \leq 6 \times 10^9\,\mathrm{u}$ the wave packet
spreads, but slower than the free solution for zero potential spreads.
The larger the mass, the slower the spreading becomes compared to the
free solution.

For masses below $10^9\,\mathrm{u}$ the difference to the free solution
is hardly identifiable.
\item For masses $m \geq 7 \times 10^9\,\mathrm{u}$ we observe a
collapsing wave packet. The shrinking behaviour was verified for
masses up to $3 \times 10^{11}\,\mathrm{u}$.
\item We have no results for larger masses, yet, as for reliable
results one needs to increase both the temporal and spatial resolution
with increasing mass, and therefore the runtime of the calculation also
increases.
\end{itemize}

\noindent In figure \ref{fig:collapse} snapshots of a typical collapse
are presented. The radial probability density
$\rho = 4 \pi \, r^2 \, \abs{\Psi}^2$ is plotted against $r$ for a mass
of $m = 7 \times 10^9\,\mathrm{u}$.

In figure \ref{fig:rp} we plot the time evolution for the global maximum
of the probability density. The collapsing behaviour for masses not less
than $7 \times 10^9\,\mathrm{u}$ is obvious. Also note the
oscillations for larger time values which are due to the packet
penetrating itself. There also seems to be some equilibrium distance to
which the wave packet collapses, which might correspond to a stationary
solution. The chaotic behaviour for the curve corresponding to
$m = 5 \times 10^{10}\,\mathrm{u}$ is probably due to numeric effects.

We also plot the radius $r_{90}$ within which 90\,\% of the probability
are located, i.\,e.
\begin{equation}
 90\,\% = \int_0^{r_{90}} 4 \pi \, r^2 \, \abs{\Psi}^2 \,\D r,
\end{equation}
against time. One can see from figure \ref{fig:r90plot}, where $r_{90}(t)$
is plotted for a mass of $10^{10}\,\mathrm{u}$, that it is
reasonable to define a collapse time by the minimum of this plot,
which then is about 30\,000 seconds.

\subsection{Coherence Time}

While interferometry experiments might still be feasible at masses of order
of the collapse mass of $10^{10}\,\mathrm{u}$ that we found, another
experimental constriction occurs. As can be seen from figures \ref{fig:rp}
and \ref{fig:r90plot} the time scale beneath which the collapse takes place
is of the order of several hours. For experimental tests one would like to take
this time scale down to a reasonable value.

Hence, we plot in figure \ref{fig:mass_time} the time it takes the solution
of the full SN-equation to differ from the solution of the free Schr\"odinger
equation in its full width at half maximum by a certain percentage against the
mass. Figure \ref{fig:mass_time}\,(a) shows that the coherence time needed
decreases with increasing mass. According to figure \ref{fig:mass_time}\,(b)
it takes, for example, about 3 minutes for the solution for
$m = 3 \times 10^{11}\,\mathrm{u}$ to differ from the free solution by one per
cent.

Consequently on one hand, if keeping the width of the wave packet constant, for an
experimental test of the SN-equation by means of e.\,g.
molecular inteferometry one needs to maximise not only the mass
but also coherence time and the sensitivity regarding the detection of
deviations from the free Schr\"odinger equation.

But note that on the other hand, we also could make use of the scaling
law \eqref{eq:SNSymm-4} to decrease the collapse time. If we simultaneously
increase the mass by factor of $\mu$ and decrease the width by a factor of
$\mu^{-3}$ the collapse time decreases by a factor of $\mu^{-5}$. Thus,
for a mass of $10^{11}\,\mathrm{u}$, for example, observed with
a grating period of 0.5\,nm one should in principle be able
to observe a collapsing wave packet causing a loss of interference
with a coherence time of approximately 300\,ms.

%%%%%%%%%%%%%%%%%%%%%%%%%%%%%%%%%%%%%%%%%%%%%%%%%%%%%%%%%%%%%%%%%%%%%%

%\bibliographystyle{plain}
%\bibliography{RELATIVITY,QM,MATH}

\begin{thebibliography}{1}

\bibitem{Arndt.Hornberger.Zeilinger:2005}
Markus Arndt, Klaus Hornberger, and Anton Zeilinger.
\newblock Probing the limits of the quantum world.
\newblock {\em Physics World}, 18:35--40, 2005.

\bibitem{Carlip:2008}
Steve Carlip.
\newblock Is quantum gravity necessary?
\newblock {\em Classical and Quantum Gravity}, 25(15):107--144, 2008.

\bibitem{Giulini:1996}
Domenico Giulini.
\newblock On {Galilei} invariance in quantum mechanics and the {Bargmann}
  superselection rule.
\newblock {\em Annals of Physics (New York)}, 249(1):222--235, 1996.

\bibitem{Hackermueller.etal:2003}
Lucia Hackerm\"uller, Stefan Uttenthaler, Klaus Hornberger, Elisabeth Reiger,
  Bj\"orn Brezger, Anton Zeilinger, and Markus Arndt.
\newblock Wave nature of biomolecules and fluorofullerenes.
\newblock {\em Physical Review Letters}, 91(9):090408 (4 pages), 2003.

\bibitem{HarrisonMorozTod:2003}
Richard Harrison, Irene Moroz, and Paul Tod.
\newblock A numerical study of the {Schr\"odinger-Newton} equations.
\newblock {\em Nonlinearity}, 16:101--122, 2003.

\bibitem{Lieb:1977}
Elliott~H. Lieb.
\newblock Existence and uniqueness of the minimizing solutions of {Choquard's}
  nonlinear equation.
\newblock {\em Studies in Applied Mathematics}, 57:93--105, 1977.

\bibitem{MorozPenroseTod:1998}
Irene Moroz, Roger Pernrose, and Paul Tod.
\newblock Spherically-symmetric solutions to the {Schr\"odinger–Newton}
  equations.
\newblock {\em Classical and Quantum Gravity}, 15(9):2733--2742, 1998.

\bibitem{RuffiniBonazzola:1969}
Remo Ruffini and Silvano Bonazzola.
\newblock Systems of self-gravitating particles in {General Relativity} and the
  concept of an equation of state.
\newblock {\em Physical Review}, 187(5):1767--1783, 1969.

\bibitem{Salzman-Carlip:2006}
Peter~Jay Salzman and Steve Carlip.
\newblock A possible experimental test of quantized gravity.
\newblock arXiv:gr-qc/0606120.
\newblock Based on the Ph.D. thesis of P.\,Salzman: ``Investigation of the Time
  Dependent Schr\"odinger-Newton Equation'', Univ. of California at Davis,
  2005.

\end{thebibliography}

\end{document}